\documentclass[12pt,preprint]{aastex}
\shortauthors{K. Itoh, M. Ishida, and H. Kunieda}
\shorttitle{Density Diagnostics of the Plasma in AE Aqr}

\newcommand{{\aaq}}{AE~Aqr}
\newcommand{{\ka}}{K$\alpha$}
\newcommand{{\ha}}{H$\alpha$}
\newcommand{{\xmm}}{\it XMM-Newton}

\begin{document}

\title{Density Diagnostics of the Hot Plasma in AE Aquarii with XMM-Newton}

\author{Kei~{\sc Itoh}\altaffilmark{1}, Manabu~{\sc Ishida}\altaffilmark{2}, 
and Hideyo~{\sc Kunieda}\altaffilmark{1}}
\email{keito@astro.isas.jaxa.jp, ishida@phys.metro-u.ac.jp,
kunieda@astro.isas.jaxa.jp}


\altaffiltext{1}{Institute of Space and Astronautical
Science, 3-1-1 Yoshinodai, Sagamihara, Kanagawa 229-8510, Japan}
\altaffiltext{2}{Department of Physics, Tokyo Metropolitan
University, 1-1 Minami-Osawa, Hachioji, Tokyo 192-0397, Japan}

\begin{abstract}
High resolution spectroscopy of AE~Aquarii with the {\xmm} RGS
has enabled us to measure the electron number density
of the X-ray-emitting hot plasma to be $\sim10^{11}$~cm$^{-3}$
by means of intensity ratios of the He-like triplet
of Nitrogen and Oxygen.
Incorporating with the emission measure evaluated by the EPIC cameras,
we have also found a linear scale of the plasma 
to be $\simeq 5\times 10^{10}$~cm.
Both these values, obtained model-independently, 
are incompatible with a standard post-shock accretion column 
of a magnetized white dwarf,
but are naturally interpreted as the plasma being formed
through interaction between an accretion flow and the magnetosphere
of the white dwarf.
Our results provide another piece of evidence 
of the magnetic propeller effect being at work in {\aaq}.

\end{abstract}


\section{Introduction}
{\aaq} is a close binary system composed of a magnetized white dwarf 
rotating at a period of $33.08$~s \citep{1979ApJ...234..978P}
and a Roche-lobe filling K3IV secondary star orbiting at a period of 9.88~h
\citep{1993ApJ...410L..39W}.
The masses of the primary and the secondary are reported to be
$M_1 = 0.79\pm 0.16M_\odot$ and $M_2 = 0.50\pm 0.10M_\odot$
\citep{1996MNRAS.282..182C}, 
and the orbital semi-major axis length is $a = 1.8\times 10^{11}$~cm.

{\aaq} has been known as one of the most enigmatic 
magnetic Cataclysmic Variables (mCVs) on various aspects, 
including large optical flares and flickering \citep{1979ApJ...234..978P},
large radio flares \citep{1996ApJ...461.1016B},
TeV $\gamma$-ray emissions \citep{1994ApJ...434..292M} and so on.
Although the orbital period of this long predicts existence
of an accretion disk,
an optical spectrum of {\aaq} does not seem as such, 
but only shows a broad single-peaked {\ha} emission
whose centroid velocity was found to lag behind the white dwarf orbit
by some 70$^\circ$ \citep{1993ApJ...410L..39W}.
Hard X-ray emissions from magnetic CVs (mCVs) 
originate from the post-shock plasma
that takes place close to the white dwarf surface in the accretion column.
Although the plasma temperature is a few tens of keV in general
\citep{1999ApJS..120..277E,1995cava.conf...93I},
that of {\aaq} is measured to be as low as $\sim 3$~keV
\citep{1999mcv..work..343E,1999ApJ...525..399C}.
All these results, together with discovery of a steady spin-down of the
white dwarf at a rate $\dot P = 5.64\times 10^{-14}$~s~s$^{-1}$
\citep{1994MNRAS.267..577D}, lead \citet{1997MNRAS.286..436W} 
to draw a picture
that the accreting matter from the secondary trapped at the magnetosphere
does not accrete onto the white dwarf 
but is expelled from the binary owing to magnetic torque 
\citep{1998MNRAS.298..285W,2001A&A...374.1030I}.
This is so-called the magnetic propeller effect.

In this paper, we present new observational results
that strongly indicate the magnetic propeller being at work in {\aaq},
by means of a model-independent density diagnostics with the He-like triplet
of Nitrogen and Oxygen fully resolved by the {\xmm} RGS for the first time.

\section{Observations}

{\aaq} was observed with {\xmm} \citep{2001A&A...365L...1J}
on 2001 November 7--8.
The observation log is given in Table \ref{tbl-1}.
\placetable{tbl-1}
The entire observation consists of two parts.
During the former $\sim$10~ks, 
both the EPIC pn \citep{2001A&A...365L..18S} and 
MOS \citep{2001A&A...365L..27T} had been switched off.
The RGS \citep{2001A&A...365L...7D}, on the other hand, 
had been normally operated throughout the observation.
As a result, 27~ks data are available for the RGS and
17~ks for the EPIC pn/MOS.
In Fig.~\ref{fig-1} shown are the light curves of pn, MOS1, and RGS1.
\placefigure{fig-1}

For the data reduction, we use the SAS version 6.0.0.
We adopt circular apertures of $54.\!\!''7$ and $50.\!\!''6$ in radius 
as the source photon integration region for pn and MOS, respectively.
The background photons are accumulated from concentric annuli,
being set out of the source regions, with three times larger outer radii.
Source photons are somewhat piled up in the pn data
especially during the flare. 
We thus have excluded the central circular region with a diameter of 16$''$
from the analysis of pn data.

\section{Data Analysis}

\subsection{The EPIC Spectra}

\begin{figure}
\epsscale{0.5}
\plotone{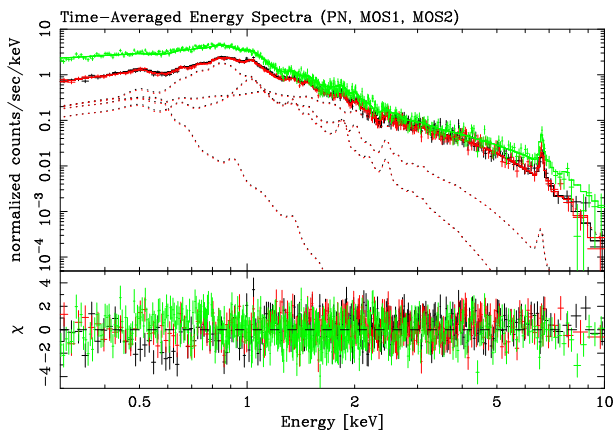}
\caption{
The averaged spectra of EPIC pn (green), MOS1 (black) and MOS2 (red)
overlaid with the best-fit four temperature {\sc vmekal} model
as the histograms.
The model components are drawn with the dotted lines for MOS1 and MOS2.
See table~\ref{tbl-2} for the best-fit parameters.
\label{fig-2}
}
\end{figure}

The time-averaged energy spectra extracted separately from the MOS and
pn data are shown in Fig.~\ref{fig-2}.
\placefigure{fig-2}
A number of K$\alpha$ emission lines of H-like
and/or He-like Ne, Mg, Si, S, and Fe can be recognized.
Coexistence of these lines indicate that the X-ray
emission originates from a multi-temperature optically thin thermal plasma
\citep{1999ApJ...525..399C,1999mcv..work..343E}.
We thus have adopted a multi-temperature {\sc vmekal} model 
\citep{1995Legacy.....6...16M} to fit these spectra.
We have added a new temperature component one by one until the addition
of a new component does not improve the fit significantly
on the basis of F-test.
To this end, we have arrived at a four temperature {\sc vmekal} model
with common elemental abundances undergoing photoelectric absorption
with a common hydrogen column density.
The best-fit parameters are listed in Table \ref{tbl-2}.
\placetable{tbl-2}
Emission measures of the four continuum components
with the temperatures of 0.14, 0.59, 1.4, and 4.6~keV are
$1.3$, $3.6$, $2.7$, and $5.3\times 10^{53}\mathrm{cm}^{-3}$,
and $1.29\times 10^{54}\mathrm{cm}^{-3}$ in total,
for the assumed distance of 100~pc \citep{1993ApJ...406..229W}.
The highest temperature $kT = 4.6$~keV is considerably lower 
than any other mCVs.
Also, the fluorescent neutral iron K$\alpha$ emission line at 6.40~keV
ubiquitous among mCVs \citep{1999ApJS..120..277E} is absent.
The upper limit of its equivalent width is obtained to be $< 88$~eV.
Thanks to a large effective area of {\xmm}, 
the abundances of the elements from N to Ni are obtained.
They are generally sub-solar \citep{1989GeCoA..53..197A}, except for N
which is more than three times the solar value.

\subsection{Line Intensities from RGS Spectra} 

\begin{figure*}
\epsscale{0.5}
\plotone{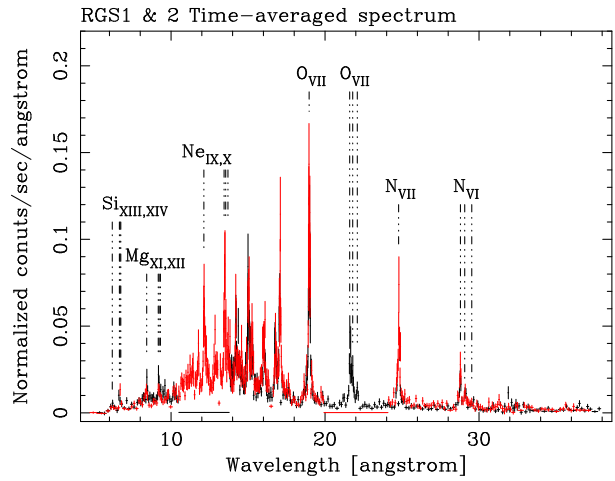}
\caption{The RGS spectra averaged over the entire 27~ks exposure.
The black and red points represent the data from RGS1 and RGS2, respectively.
Identifications of the H-line and He-like K$\alpha$ lines of N through Si
are shown with dotted-dashed line.
The other unidentified lines are mainly those associated with Fe-L transitions.
\label{fig-3}
}
\end{figure*}

\begin{figure}
\epsscale{0.8}
\plotone{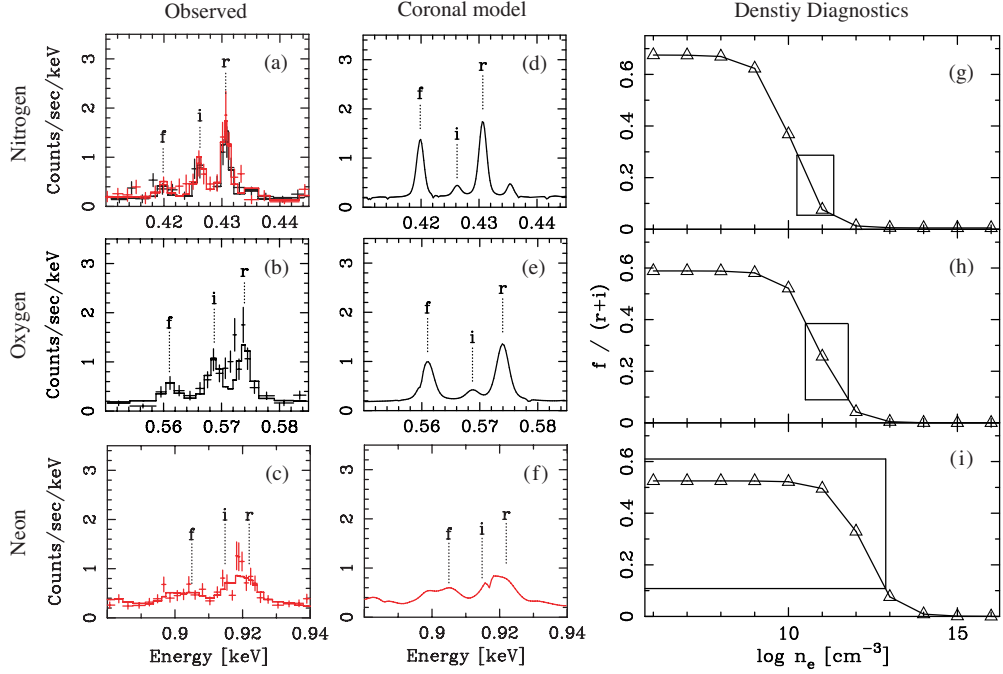}
\caption{Density diagnostics by means of the He-like triplet.
Panels of observed and coronal limit spectra 
around the He-like triplet, and the density diagnostics
are arranged from left to right columns.
The cases of Nitrogen, Oxygen, and Neon are arranged from upper to lower raws.
Black and red data points and histograms are from RGS1 and RGS2.
The relative intensities of the intercombination (marked with `i')
and forbidden (marked with `f') lines are undoubtedly inverted for N and O.
In the panels (g)-(i), 
comparisons of theoretical curves of the intensity ratio $f/(r+i)$
versus the electron density $n_{\rm e}$ with the observed intensity ratio.
In drawing the theoretical curves we have used 
the plasma code {\sc spec} \citep{1996uxsa.coll..411K}
at the ionization temperatures $kT_i$ of 0.15, 0.30, and 0.34~keV
for N, O, and Ne obtained from the intensity ratio between Ly$\alpha$ and $r$.
\label{fig-4}
}
\end{figure}

In Fig.~\ref{fig-3} we have shown the RGS spectra.
\placefigure{fig-3}
The K$\alpha$ emission lines from Nitrogen through Silicon
in the H-like and/or He-like ionization states can easily be recognized.
Figure~\ref{fig-4} (a)-(c) are blow-up of energy bands of 
the He-like triplets of N, O, and Ne.
\placefigure{fig-4}
In the next column shown are the spectra predicted 
by the best-fit four temperature {\sc vmekal} model (table~\ref{tbl-2}),
being convolved by the RGS1 1st order response function.
Note that the {\sc vmekal} model represents emission spectrum
from a thermal plasma in the low density coronal limit.
One can easily see, for N and O, that the intensity of the
intercombination increases from the coronal limit
by consumption of the forbidden.
This behavior is interpreted as a high density effect;
if the electron density exceeds a certain critical value inherent in
each element, one of the two electrons excited to the upper level 
of the forbidden line ($^3\mathrm{S}_1$) is further pumped by another impact
of a free electron up to the higher level $^3\mathrm{P}_{2,\,1}$,
and is then relaxed by radiating the intercombination line.
The relative intensities of the intercombination and forbidden lines
can therefore be utilized as a density diagnostics
\citep{1969MNRAS.145..241G,1981ApJ...249..821P}.

In order to evaluate the electron number density $n_\mathrm{e}$,
we begin with evaluating intensities of the He-like triplets.
For this, we utilize the four temperature {\sc vmekal} model
that provides the best-fit to the EPIC pn and MOS spectra.
We have fixed the hydrogen column density and the four temperature 
at the values in table~\ref{tbl-2}
and the relative normalizations of the four continuum components.
The abundances are also fixed at the best-fit values, except for an element
to be used as a density diagnostic, 
for which the abundance is set equal to zero, and instead,
four Gaussians are added, 
representing the He-like triplet and the Ly$\alpha$ line.
The line intensities thus obtained for N, O, and Ne are summarized in
table~\ref{tbl-3}, and the best-fit results are displayed in Fig.~\ref{fig-4}
(a)-(c) as the histograms.
\placetable{tbl-3}
The ionization temperature $kT_i$
calculated from the intensity ratio between the Ly$\alpha$ and $r$
are obtained to be 0.16, 0.30, and 0.34~keV for N, O, and Ne,
respectively.

Given the line intensities of the triplets,
we have carried out density diagnostics by means of the intensity ratio
$f/(r+i)$.
In Fig.~\ref{fig-4} (g)-(i) shown are theoretical curves
of the ratio $f/(r+i)$ versus the electron density,
which is calculated by means of the plasma code {\sc spex} 
\citep{1996uxsa.coll..411K} at the ionization temperature $kT_i$
of each element.
In each panel we have also drawn a range of the intensity ratio
allowed from the data and the resultant density range as a box.
The electron densities are obtained
to be $1.4\times 10^{10}-1.3\times 10^{11}\mathrm{cm}^{-2}$,
$4.0\times 10^{10}-6.8\times 10^{11}\mathrm{cm}^{-2}$,
and $< 9.3\times 10^{12}\mathrm{cm}^{-2}$ for N, O, and Ne, respectively.
As a crude approximation, the electron number density
of the plasma is $n_{\rm e} \simeq 10^{11}{\rm cm}^{3}$.

\section{Discussion and Conclusion}

It is of great importance to note that the resultant density
$n_{\rm e} \simeq 10^{11}{\rm cm}^{3}$ is smaller 
by several orders of magnitude 
than the conventional estimate in the post-shock accretion column of mCVs: 
$n_{\rm e}\simeq 10^{16}{\rm cm}^{-3}$ \citep{2002apa..book.....F}.
Moreover, combined with the emission measure 
$EM = 1.3\times 10^{54}{\rm cm}^{-3}$ obtained with the EPIC pn/MOS (\S~3.1),
the linear scale of the plasma is evaluated to be 
$\ell_\mathrm{p} = (EM/n_{\rm e}^2)^{1/3} \simeq 5\times 10^{10}$~cm,
which is much larger than a radius of a typical white dwarf
and rather close to the orbital scale.
We thus conclude that the optically thin hot plasma in {\aaq} does not
accrete onto the white dwarf but rather spreads over the orbit of the binary.
Note that this conclusion is derived simply
on the basis of widely approved atomic physics, and hence, 
is free from any specific model.

It is known, however, that photo-excitation due to UV radiation can
also pump $^3\mathrm{S}_1$ electron up to $^3{\rm P}_{2,\,1}$,
thereby affecting the density diagnostics.
Although there is no evidence of accretion disk in {\aaq} 
\citep{1993ApJ...410L..39W} and the secondary star is a late type K3 subgiant,
the white dwarf can be a source of the photo-excitation, 
the emission spectrum of which is reported to be a blackbody
with a temperature of $T_{\rm r} = 26,000$~K \citep{1999mcv..work..357W}.
The photo-excitation rate of such a white dwarf 
from the initial state $i =\; ^3\mathrm{S}_1$
to the final state $j =\; ^3{\rm P}_{2,\,1}$ is
\begin{equation}
\Gamma_{ij}(T_{\rm r}) \;=\; \frac{\pi e^2}{m_{\rm e}c}\,f_{ij}\cdot W\cdot
    \frac{\pi B_\nu (T_{\rm r})}{h\nu}
\end{equation}
\citep{2001A&A...376.1113P,2002pcvr.conf..113M} 
where $f_{ij}$ is the effective oscillator strength of photo-excitation
which is 0.03574 \citep{1999A&AS..135..347N},
and $W$ is a geometrical dilution factor to be taken here as
$(R_{wd}/\ell_{\rm p})^2$ where $R_{wd}$ is the radius of the white dwarf.
On the other hand, 
the collisional excitation rate to be compared with $\Gamma_{ij}$ is
$n_{\rm e} q_{ij}$, where,
\begin{equation}
q_{ij}(T_{\rm e})\;=\;
  \frac{8.63\times 10^{-6}}{\omega_i\,T_{\rm e}^{1/2}}\,
    e^{-E_{ij}/kT_{\rm e}}\gamma_{ij}(T_{\rm e})\;\;\;{\rm [\,cm^3\;s^{-1}\,]}
\label{eq:colrate}
\end{equation}
is the rate coefficient of the electron impact excitation
\citep{1978A&A....65...99M}.
In this equation, $\omega_i (= 3)$ is the statistical weight of the
lower level, $E_{ij}$ is the energy difference between the two levels,
and $\gamma_{ij}$ is the Maxwellian-averaged collision strength
which is tabulated as a function of $T_{\rm e}$ in \citet{1981ApJ...246.1031P}.
Adopting $T_i$ listed in table~\ref{tbl-3} as $T_{\rm e}$,
we compared eqs.~(1) and (2) and obtained 
$\Gamma_{ij} \simeq 0.07\,n_{\rm e} q_{ij}$ for both N and O.
Hence, the photo-excitation effect can be neglected
in the first order approximation.

We remark that the density obtained in \S~3.2 should be treated 
as an upper limit, in the case that the UV radiation is stronger 
than the estimation above.
Even if so, the conclusion above need not to be changed because
a lower density, and hence, a larger geometrical scale are required
for the plasma.

The low density and the large scale of the plasma naturally lead us to invoke
the magnetic propeller model, in which blobby accreting matter 
originally following a ballistic trajectory from the inner Lagrangian point
is gradually penetrated by the magnetic field of
the white dwarf, and is finally blown out of the binary
due to interaction with the rapid rotating white dwarf magnetosphere
\citep{1997MNRAS.286..436W}.
The magnetic propeller model is advantageous in explaining 
various characteristics of {\aaq}, such as 
the spin down of the white dwarf, the velocity modulation of H$\alpha$ line.
The low plasma temperature compared with other mCVs
can be attributed to a halfway release of the gravitational potential energy
of accreting matter down to the magnetosphere.
We believe that
the density diagnostics presented here add another piece of
evidence to support the magnetic propeller effect being at work in {\aaq}.

\begin{figure}
\epsscale{0.5}
\plotone{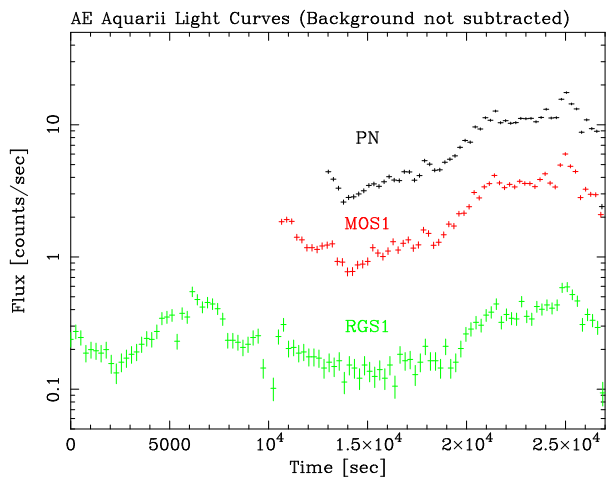}
\caption{
The light curves of EPIC pn and MOS1 in the band 0.2--15~keV, 
and RGS1 in the band 5--35~\AA with a bin size of 256~sec.
The source integration region is a circle with a radius of
$57.\!\!''4$ and $50.\!\!''6$ for pn and MOS1, respectively.
The RGS1 light curve is created with the 1st order photons.
\label{fig-1}
}
\end{figure}

\citet{1998MNRAS.298..285W} estimated that the velocity of the out-flowing
plasma can be as high as several hundred km at most.
If so, a Doppler shift of the N and O lines may be detected through
phase-resolved spectral analysis.
We thus have made RGS spectra of the first and second flares separately,
during the time interval of 3,000-8,500~s and 19,000-26,000~s 
in Fig.~\ref{fig-1}, respectively,
and evaluated the central energy of Oxygen Ly$\alpha$.
The result is negative with the line central energies of
$653.4\pm 0.5$~eV and $653.4\pm 0.4$~eV, respectively,
which are fully consistent with the laboratory value.
We have further split the data of the second flare into evenly segregated
three pieces, but the central energy distributes in the range
653.2-653.4~eV.
Note, however, that we have obtained a finite value 
for a 1-$\sigma$ line width of $2.4\pm 0.6$~eV and $1.7\pm 0.6$~eV 
for the first and second flares, respectively, 
which are larger than the natural width ($\sim 0.001$~eV)
or the thermal velocity of Oxygen ion ($\sim 70$~km~s$^{-1}$ or
$\sim 0.2$~eV estimated from $kT_i = 0.3$~keV).
Since the ionization temperature of Oxygen is lower than the highest
temperature of the plasma (table~\ref{tbl-2}) by an order of magnitude, 
and also from the fact $\ell_{\rm p} \simeq 5\times 10^{10}$~cm, 
the Oxygen Ly$\alpha$ line emanates from a region
far out of the magnetosphere where the velocity collimation is
already dissolved.

In order to finally confirm the magnetic propeller effect being 
at work in {\aaq},
it is important to detect a bulk velocity shift of an emission line
expected from the plasma flow in the vicinity of the magnetosphere.
Since the highest temperature of $\sim 5$~keV, 
we expect this can be done by the iron K$\alpha$ line
with the {\it Astro-E2} XRS.

\acknowledgments

\begin{deluxetable}{cccllc}
\footnotesize
\tablecaption{Observation Log \label{tbl-1}}
\tablewidth{0pt}
\tablehead{
\colhead{Obs. ID} & \colhead{Instrument} & \colhead{Data Mode} 
  & \colhead{Obs. Start (UT)} & \colhead{Obs. End (UT)} & \colhead{Exp.(s)}
}
\startdata
   0111180601& RGS1 & Spectroscopy & 2001-11-07 20:06:35 &
   2001-11-07 22:53:33 & 10018 \\
             & RGS2 & Spectroscopy & 2001-11-07 20:06:35 &
   2001-11-07 22:53:28 & 10013 \\ \hline
   0111180201& MOS1 & Large Window & 2001-11-07 23:06:36 &
   2001-11-08 03:42:09 & 16533 \\
             & MOS2 & Large Window & 2001-11-07 23:06:35 &
   2001-11-08 03:42:08 & 16533 \\
             & pn & Full Frame & 2001-11-07 23:45:53 &
   2001-11-08 03:38:17 & 13910 \\
             & RGS1 & Spectroscopy & 2001-11-07 23:00:19 &
   2001-11-08 03:45:39 & 17120 \\
             & RGS2 & Spectroscopy & 2001-11-07 23:00:19 &
   2001-11-08 03:45:36 & 17117
\enddata
\end{deluxetable}

\begin{deluxetable}{cccc}
\footnotesize
\tablecaption{The best-fit parameters of the 4 temperature vmekal model
being fit to EPIC pn and MOS spectra \label{tbl-2}
}
\tablewidth{0pt}
\tablehead{
\colhead{Parameter} & \colhead{Value}   & \colhead{Element}   
	&  \colhead{Abundance\tablenotemark{a}}
} 
\startdata
   $N_{\rm H}$ $(10^{20} \mathrm{cm^{-2}})$ & $3.59^{+1.47}_{-1.20}$
	& N & $3.51^{+0.92}_{-0.81}$ \\
   $kT_{1}~(\mathrm{keV})$ & $4.60^{+0.60}_{-0.47}$ 
	& O & $0.74^{+0.17}_{-0.23}$ \\
   $kT_{2}~(\mathrm{keV})$ & $1.21^{+0.13}_{-0.08}$ 
	& Ne & $0.43^{+0.28}_{-0.25}$ \\
   $kT_{3}~(\mathrm{keV})$ & $0.59^{+0.02}_{-0.02}$ 
	& Mg & $0.70^{+0.15}_{-0.14}$ \\
   $kT_{4}~(\mathrm{keV})$ & $0.14^{+0.05}_{-0.02}$ 
	& Si & $0.81^{+0.15}_{-0.12}$ \\
   $N_1$\tablenotemark{b} $(10^{-3})$ & $4.45^{+0.41}_{-0.44}$ 
	& S & $0.73^{+0.20}_{-0.18}$ \\
   $N_2$\tablenotemark{b} $(10^{-3})$ & $2.25^{+0.52}_{-0.51}$
	& Ar & $0.21~(<0.89)$ \\
   $N_3$\tablenotemark{b} $(10^{-3})$ & $3.04^{+0.47}_{-0.41}$
	& Ca & $0.19~(<1.11)$ \\
   $N_4$\tablenotemark{b}~$(10^{-3})$ & $1.07^{+0.64}_{-0.37}$
	& Fe & $0.47^{+0.07}_{-0.06}$ \\
   $N_{\rm pn}/N_{\rm MOS}$\tablenotemark{c} & $1.08^{+0.01}_{-0.01}$ 
	& Ni & $1.27^{+0.57}_{-0.50}$ \\
   $\chi^2_\nu$ (d.o.f.) & 1.22 (992) & &
\tablenotetext{a}{Solar Abundances \citep{1989GeCoA..53..197A}}
\tablenotetext{b}{Normalization of the {\sc vmekal} component obtained with
pn camera in a unit of $10^{-14}/4 \pi D^2 \int n_e n_H~dV$,
where $D~[\mathrm{cm}]$ is the distance to the target star.}
\tablenotetext{c}{Ratio of continuum normalizations.}
\enddata
\tablecomments{All the errors are at the 90~\% confidence level.}
 
\end{deluxetable}

\begin{deluxetable}{ccccccc}
\footnotesize
\tablecaption{
Intensities of Ly$\alpha$ and He-like triplet 
of Nitrogen, Oxygen and Neon with RGS \label{tbl-3}
}
\tablewidth{0pt}
\tablehead{
 \colhead{}  &\multicolumn{2}{c}{Nitrogen}  &\multicolumn{2}{c}{Oxygen} 
   & \multicolumn{2}{c}{Neon} \\
\cline{2-3} \cline{4-5} \cline{6-7} 
 \colhead{} 
   & \colhead{Energy\tablenotemark{a}} & \colhead{Norm\tablenotemark{b}}
   & \colhead{Energy\tablenotemark{a}} & \colhead{Norm\tablenotemark{b}}
   & \colhead{Energy\tablenotemark{a}} & \colhead{Norm\tablenotemark{b}}
}
\startdata
 Ly$\alpha$ & 0.50032 & $1.85^{+0.21}_{-0.21}$ & 0.65348 
   & $3.63^{+0.34}_{-0.35}$ & 1.0215 & $ 0.90^{+0.23}_{-0.23}$ \\
 $r$        & 0.43065 & $0.97^{+0.17}_{-0.17}$ & 0.57395 
   & $1.06^{+0.22}_{-0.23}$ & 0.92195 & $0.65^{+0.22}_{-0.22}$ \\ 
 $i$        & 0.42621 & $0.57^{+0.17}_{-0.17}$ & 0.56874 
   & $0.85^{+0.22}_{-0.22}$ & 0.91481 & $0.24^{+0.21}_{-0.21}$ \\ 
 $f$ & 0.41986 & $0.26^{+0.14}_{-0.13}$ & 0.56101 
   & $0.37^{+0.17}_{-0.18}$ & 0.90499 & $0.32^{+0.18}_{-0.18}$ \\
\cline{1-7}
 $kT_{i}$\tablenotemark{c}  & \multicolumn{2}{c}{$0.16\pm 0.02$}
   & \multicolumn{2}{c}{$0.30^{+0.04}_{-0.03}$} 
      & \multicolumn{2}{c}{$0.34^{+0.09}_{-0.07}$} \\
 $\chi^{2}_{\nu}$ (d.o.f.)& \multicolumn{2}{c}{1.36 (115)} 
   & \multicolumn{2}{c}{1.14 (69)} & \multicolumn{2}{c}{1.21 (63)}
\tablenotetext{a}{Line central energy in keV.}
\tablenotetext{b}{Line normalization 
in a unit of $10^{-4}$~photons~cm$^{-2}$~s$^{-1}$.}
\tablenotetext{c}{Ionization temperature in a unit of keV, 
evaluated by the intensity ratio between Ly$\alpha$ and $r$.}
\enddata
\tablecomments{All the errors are at the 90~\% confidence level.}
\end{deluxetable}

\end{document}